\documentclass[epj]{svjour}
\usepackage{amsmath}
\usepackage[english]{babel}
\usepackage{graphicx}
\usepackage{epstopdf}
\usepackage{subfig}

\usepackage{booktabs}
\usepackage{fancyref}
\usepackage[numbers,sort&compress]{natbib}
\usepackage[separate-uncertainty=true]{siunitx}
\usepackage[bookmarks=false]{hyperref}
\hypersetup{
	colorlinks   = true,    
	urlcolor     = blue,    
	linkcolor    = blue,    
	citecolor    = blue     
}
\usepackage{caption}
\begin{document}


\title{Modeling of nanoparticle coatings for medical applications}
\author{Kaspar Haume \inst{1,2,}%
	\thanks{\email{kaspar.haume@open.ac.uk}}%
	\and Nigel J. Mason \inst{1}%
	\and Andrey Solov'yov \inst{2,3}%
	\thanks{On leave from A. F. Ioffe Physical Technical Institute, St. Petersburg 194021, Russian Federation}%
}                     
\institute{
	Department of Physical Sciences, The Open University, Milton Keynes MK7 6AA, UK 
	\and MBN Research Center, Altenh\"oferallee 3, 60438 Frankfurt am Main, Germany 
}
\date{Received: date / Revised version: date}
%

\abstract{
	Gold nanoparticles (AuNPs) have been shown to possess properties beneficial for the treatment of cancerous tumors by acting as radiosensitizers for both photon and ion radiation. Blood circulation time is usually increased by coating the AuNPs with poly(ethylene glycol) (PEG) ligands.
	The effectiveness of the PEG coating, however, depends on both the ligand surface density and length of the PEG molecules, making it important to understand the structure of the coating. In this paper the thickness, ligand surface density, and density of the PEG coating is studied with classical molecular dynamics using the software package MBN Explorer.
	AuNPs consisting of 135 atoms (approximately \SI{1.4}{nm} diameter) in a water medium have been studied with the number of PEG ligands varying between 32 and 60.
	We find that the thickness of the coating is only weakly dependent on the surface ligand density and that the degree of water penetration increased when there is a smaller number of attached ligands.
}

\maketitle
%


\section{Introduction\label{sec:Introduction}}
Radiotherapy with x-rays or gamma rays is a widespread methodology to treat cancer tumors. However, due to the efficient penetration of tissue by these photons, a considerable fraction of the total dose is deposited in healthy tissue before and after the tumor leading to potentially severe side-effects. In recent years several studies have demonstrated the radiosensitizing effect of metal nanoparticles (NPs) leading to a higher therapeutic index (ratio of therapeutic efficacy to side effects) \citep{Hainfeld2004, Polf2011, Porcel2012, Sancey2014b}. Dose localization by use of NPs has become a subject of significant scientific interest in the last decade, in part due to the promises of fewer side-effects for cancer patients worldwide, but also due to the exciting interdisciplinary nature involving biology, atomic cluster physics, collision studies, materials engineering. A core component of this research are computational efforts to model the interactions between radiation, NPs, and biological matter.

It is widely accepted that the main cell killing pathway during cancer radiotherapy is mediated by secondary electrons and radicals \citep{Porcel2012, Carter2007, Usami2008, Surdutovich2014}. The sensitizing effect of metal NPs is related to an increased emission of secondary electrons compared to a similar volume of water \citep{Verkhovtsev2015a}. These electrons in turn activate hydrolysis of the surrounding water medium resulting in an increased overall radical yield. For this reason, much effort is currently devoted to understanding and predicting the capabilities of NPs to emit secondary electrons. High-Z elements (high atomic number), such as noble metals, are particularly efficient Auger electron emitters and have been shown to generate radiosensitization through increased radical yield \citep{Kobayashi2010, Coulter2013, Porcel2010}. 

Gold nanoparticles (AuNPs), especially, have become a popular choice since the first demonstration of their radiosensitization properties \citep{Hainfeld2004}. In addition a high interaction cross section with photon radiation, their biological inertness, established methods of synthesis in a wide range of sizes and shapes, and possibility to coat their surface with a large catalog of molecules, providing the ability to partially control the behavior of the AuNPs, make them an attractive choice \citep{Hainfeld2008,McMahon2011,Kwatra2013}.  

NPs are unstable in physiological conditions and tend to agglomerate and to be eliminated from the bloodstream \citep{Alexis2008}. For this reason, AuNPs are usually coated with the molecule poly(ethylene glycol) (PEG), a process known as PEGylation, which has been shown to increase blood circulation time (time before the NP is eliminated from the bloodstream) and improve stability (reduce tendency for NPs to aggregate) \citep{Otsuka2003, Jokerst2011, Walkey2012}. In the scenario of radiosensitization, however, the effect of the coating is not clear. Although radiosensitization with PEGylated AuNPs has been demonstrated \citep{Liu2010, Zhang2012}, Gilles et al. showed that the hydroxyl radical yield was diminished for AuNPs coated with PEG depending on the coating density \citep{Gilles2014}. In another study, Xiao et al.\ found a decrease in sensitization through secondary electrons for increasing coating thickness \citep{Xiao2011}.

A better understanding of the structure and dynamics of the coating of AuNPs is therefore necessary to be able to predict their radiosensitizing properties as well as their interaction with the environment. The sizable number of possible coating molecules, including antibodies, proteins, sugars, and other organic compounds such as acids, make for a vast landscape of core-coating combinations. Experimentally investigating all possible combinations in a systematic manner is a staggering task. In this paper, we take an alternative approach by using computer simulations to study a specific combination of coating and NP core, namely the PEGylated AuNP. The presented method is general and is not restricted to the systems considered here thereby providing a convenient framework to study any core-coating combination.

Specifically in this paper classical molecular dynamics is used to simulate AuNPs of 135 atoms (approximately \SI{1.4}{nm} diameter) coated with between 32 and 60 thiolated PEG-amine ($\mathrm{S-PEG_5-NH_2}$) ligands. The system is fully solvated with water molecules. Using MBN Explorer \citep{Solovyov2012} we report the effect of coating ligand density on the coating layer thickness and density.

The paper is structured as follows: After this introduction, the computational details of the simulations are presented, divided into the preparation of the metal core, the preparation of the coating molecules and the solvation of the system, and finally the details of the molecular dynamics simulations. This section is followed by a presentation of the results and a discussion before ending with a summarizing conclusion.


\section{Methodology}
\subsection{Preparation of metal core}
\begin{figure}[h]
	\begin{centering}
		\includegraphics[width=0.80\columnwidth]{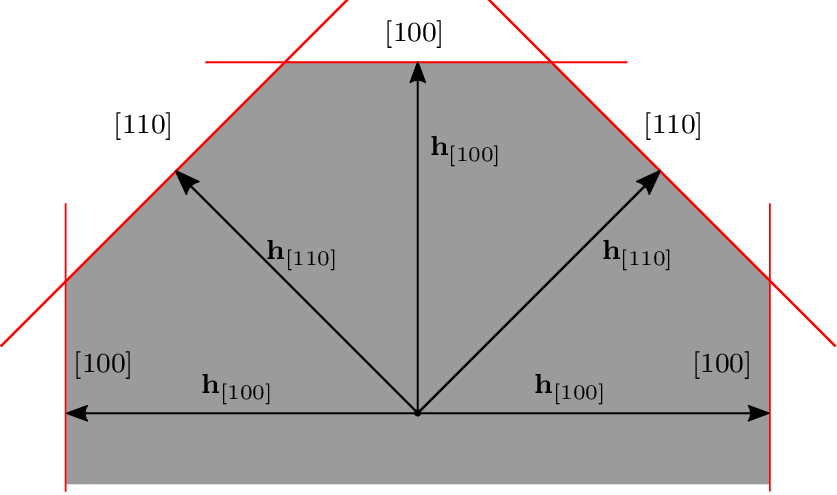}
	\par\end{centering}
	\caption{\label{fig:wulff}Wulff construction in 2D (only upper half shown). The nanoparticle
		(gray area) will be the smallest area enclosed by the red lines,
		see text}
\end{figure}

The AuNP core was created using the Wulff construction plugin of the software Virtual NanoLab\footnote{http://www.quantumwise.com} (version 2015.1). The Wulff construction is a simple theoretical approach in two steps to approximate the shape of a nanosized crystal (e.g. a NP) based on the surface energy of the faces of the crystal. In the first step a vector  $\mathbf{h}_{j}$ is drawn from the center of the NP normal to each of the crystal faces $j$ relevant for the given material --- see Fig.~\ref{fig:wulff} for a 2D example. In the second step a line is drawn perpendicularly to each vector at the end of them, the NP will then be the internal volume enclosed by these lines, similar to how the Wigner-Seitz cell is constructed. The shape of a real NP is determined by the surface energies of the crystal faces $\gamma_{j}$. This enters into the Wulff construction by setting the length of each vector proportional to the surface energy of the given face: $\left|\mathbf{h}_{j}\right|\propto\lambda\gamma_{j}$, where $\lambda$ is a constant which can be chosen and acts to scale the overall volume of the NP. A small surface energy will then lead to a larger crystal face.

To model the interactions between the Au atoms of the NP core the Sutton-Chen potential was used with parameters taken from \citep{Sutton1990} and a cutoff of \SI{5.8}{\AA} corresponding to the third-nearest neighbor for bulk gold in the face-centered cubic crystal structure. The Sutton-Chen potential has the form 
\begin{equation}
	U_{\mathrm{tot}}=\epsilon\sum_{i}
		\left[
			\frac{1}{2} \sum_{j\neq i} \left( \frac{a} {r_{ij}} \right)^{n} - 
			c\sqrt{\rho_{i}}
		\right],
\end{equation}
where 
\begin{equation}
	\rho_{i}=\sum_{j\neq i} \left( \frac{a} {r_{ij}} \right)^{m},
\end{equation}
$r_{ij}$ is the separation between the $i$'th and the $j$'th atom, $c$ is a dimensionless parameter, $\epsilon$ is a parameter with the dimensions of energy, $a$ is the bulk lattice constant of the metal, and $m$ and $n$ are positive integers with $n>m$. For Au $m=8$, $n=10$, $c=34.408$, $a=\SI{4.07825}{\AA}$, and $\epsilon=\SI{1.2793e-2}{eV}$.

To get a better starting point for the AuNP core before attaching the PEG ligands the AuNP was annealed in vacuum by first thermalizing for \SI{50}{ps} ramping the temperature from \SI{0}{K} to \SI{300}{K} with an integration time step of \SI{1}{fs} and temperature control provided by the Langevin thermostat with a time constant of \SI{0.2}{ps} \citep{Frenkel2001}. These time constants were  used for all the following MD simulations. Subsequently, the system was heated to \SI{1400}{K} for a total of \SI{400}{ps}. The NP was then cooled down to \SI{0}{K} in steps of \SI{100}{K}, each lasting \SI{50}{ps}. 

The potential energy of the annealed AuNP was compared to data on globally optimized metal clusters from the Cambridge Cluster Database\footnote{http://www-wales.ch.cam.ac.uk/CCD.html} and was shown to be nearly identical indicating that it is a good starting point on which to apply the PEG coating. It should be noted that in the present study we were not interested in global minimum configurations. The following annealing procedures will cause reorganization of the atoms and the investigated parameters (density, thickness) of the system will be evaluated at finite temperatures.

\subsection{Preparation of PEG coating and solvation}
To obtain the files necessary to describe the PEG molecule, Marvin Sketch\footnote{Version 15.4.27.0, 2015, ChemAxon (http://www.chemaxon.com)} was used to draw it. The resulting MOL2 structure file was uploaded to the SwissParam server\footnote{http://www.swissparam.ch/} to obtain the PDB structure file \citep{Zoete2011a}, and to the CHARMM General Force Field (CGenFF) site\footnote{http://cgenff.paramchem.org/} to obtain the topology and parameter files for use with the CHARMM force field \citep{Vanommeslaeghe2010a}. Finally, the protein structure file (PSF) generation plugin tool (version 1.2) within VMD (version 1.9.1) \citep{Humphrey1996} was used to generate the PSF file and a new PDB file to ensure a proper format of the files.

\begin{figure}[h]
	\begin{centering}
		\subfloat{
			\includegraphics[width=0.3\columnwidth]{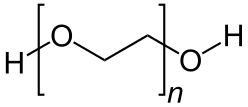}
		}
		\quad
		\subfloat{
			\includegraphics[width=0.6\columnwidth]{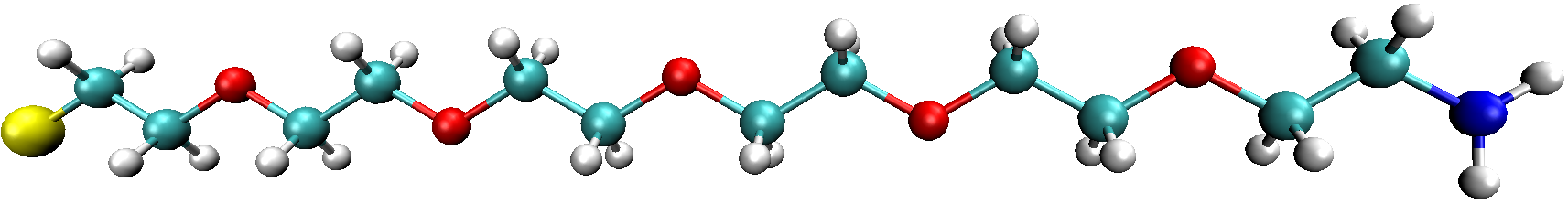}
		}
		\par\end{centering}
	\caption{(a) The chemical structure of the PEG molecule. (b) The thiolated PEG-amine molecule considered in this paper. Yellow: sulfur, teal: carbon, white: hydrogen, red: oxygen, blue: nitrogen.}
	\label{fig:PEGmolecule}
\end{figure}

It is currently accepted that the sulfur-passivating hydrogen atom of the thiol group dissociates upon bond formation with gold \citep{Kankate2009,Barngrover2013}. However, the CHARMM force fields used in these simulations do not allow bond formation and breaking. This was overcome by manually removing the hydrogen atom and applying its partial charge $q_{\mathrm{H}}$ evenly to the gold atoms of the AuNP such that every gold atom of the NP was assigned a partial charge of $q_{\mathrm{Au}}=N_{\mathrm{PEG}}q_{\mathrm{H}}/N_{\mathrm{Au}}$ where $N_{\mathrm{PEG}}$ is the number of attached PEG molecules and $N_{\mathrm{Au}}$ is the number of gold atoms in the AuNP. See Fig.~\ref{fig:PEGmolecule} for an illustration of the general PEG molecule and the one used here.

Experimental and theoretical estimates of the coverage of AuNPs by alkanethiols provide a starting point around which we can decide on a range of $N_{\mathrm{PEG}}$ to simulate. The number of surface atoms of a NP scales with the number of atoms in the sphere to the power 2/3. Since ligands form bonds mainly with surface atoms, the number of attached ligands should therefore, as a first approximation, scale with the number of NP atoms to the power 2/3. Dass estimated, from a review of experimental measurements of thiolated AuNPs, this proportionality factor and found that the number of attached ligands $N_L$ scales as $N_L = cN_{\mathrm{Au}}^{2/3}$, where $c = 1.82 \pm 0.33$ and $N_{\mathrm{Au}}$ is the number of atoms in the gold core \citep{Dass2012}. In the present paper, AuNPs of 135 atoms are simulated which gives lower and upper bounds of $N_L = 39$ and 57, respectively. This is supported experimentally by Badia et al. who found a ligand surface density of $\SI{5.9}{nm^{-2}}$ and $\SI{6.7}{nm^{-2}}$ for $\mathrm{C_{14}SH}$ and $\mathrm{C_{18}SH}$ coated AuNPs of diameter between 20 and \SI{30}{A} assuming spherical NPs \citep{Badia1996}. These footprints equate to \num{35.8+-0.8} and \num{40.5+-1.1} ligands, respectively, assuming a spherical NP of \SI{1.4}{nm} diameter, as in the present study. Theoretically, Djebali et al. found a similar ligand surface density of \SI{6.3}{nm^{-2}} for alkanethiol coated icosahedral AuNPs of \SI{2}{nm} diameter \citep{Djebaili2013}. 

Based on these numbers, we decided to study PEGylated AuNPs with $N_\mathrm{PEG}$ between 32 and 60 molecules evenly spread out approximating the NP as a spherical particle.

The Au-S interaction was modeled by a Lennard-Jones potential as given by 
\begin{equation}
	U_{\mathrm{tot}} = \epsilon\sum_{i<j}^{N}
		\left[
			\left(  \frac{r_{\mathrm{min}}} {r_{ij}} \right)^{12} - 
			2\left( \frac{r_{\mathrm{min}}} {r_{ij}} \right)^{6} 
		\right] 
\end{equation}
with minimum-energy interatomic separation $r_{\mathrm{min}}=\SI{3.0}{A}$ and the potential well depth $\epsilon=\SI{3}{eV}$. The minimum potential separation $r_{\mathrm{min}}$ was taken from data from \citep{Wright2013}.

The use of the Lennard-Jones potential for the Au-S interaction is a rather crude approximation. The exact interaction of the sulfur atoms on the gold surface is not the  focus of this research, however, and the interaction and possible intercalation of sulfur atoms into the surface of the AuNP is deemed of little importance for the overall structure of the organic part of the coating. The Au-S bond is, in itself, a topic of intense research due to its surprisingly complicated nature and is best studied with density functional theory or quantum chemistry, see for example the references by Mariscal et al. \citep{Mariscal2010} or Malola and H\"{a}kkinen \citep{Malola2015} for more information.

Finally, the PEGylated AuNP was solvated with water using the solvate plugin (version 1.6) of VMD with a water padding of \SI{20}{\AA} on all sides. The TIP3P water model is used for interactions between the water molecules \citep{Jorgensen1983}.

The initial system is shown in Fig.~\ref{fig:32PEGwaterbox-initial}, where the waterbox dimensions have been reduced for illustration purposes.

\subsection{MD simulations }
For all molecular dynamics simulations, MBN Explorer (version 2.0) was used \citep{Solovyov2012} with the following procedure. Each system was first optimized using the velocity quenching algorithm of MBN Explorer for \num{20000} steps to avoid overlapping atoms. The optimization was followed by an equilibration simulation to get the correct density of water. This was done by applying a sufficiently thick vacuum padding around the system such that it was essentially isolated from its periodic images and free to expand or contract. The \SI{400}{ps} duration of the equilibration was enough to obtain a constant density of water. Due to the tendency of water in vacuum to form a droplet, a rectangular box was cut out of the resulting system for further use. To ensure a negligible interaction between the coating and its periodic images during the MD simulations, the side lengths of the box were chosen such that there would be a layer of water of at least \SI{10}{\AA} around the coating resulting in a cubic system of side length approximately \SI{8}{nm}. Subsequently, a new optimization was performed for \num{20000} steps to eliminate any overlapping atoms. These equilibration simulations were done at \SI{310}{K} with a time step of \SI{1}{fs} and with temperature control provided by the Langevin thermostat with a time constant of \SI{0.2}{ps}.

\begin{figure}[h]
	\begin{centering}
		\subfloat[\label{fig:32PEGwaterbox-initial}] {
			\includegraphics[width=0.3\textwidth]{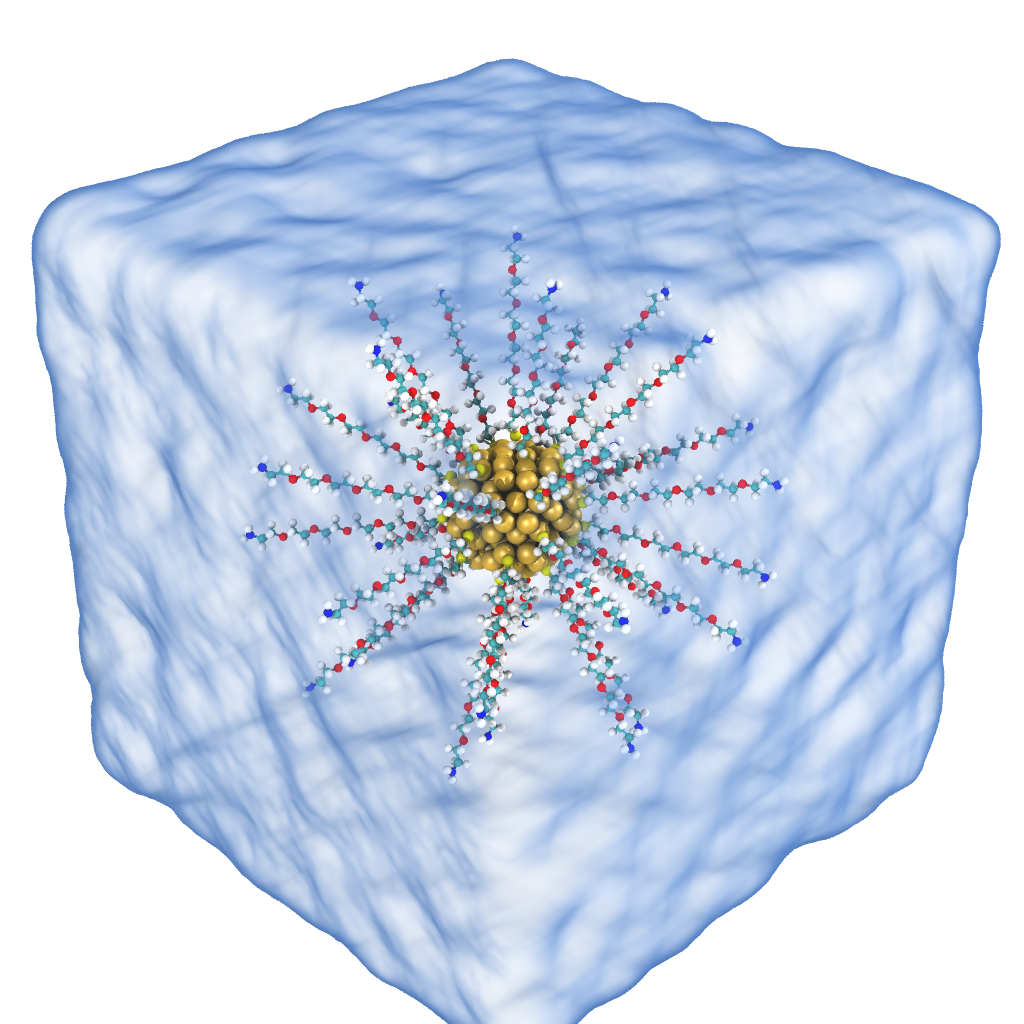}
		} \qquad
		\subfloat[] {
			\includegraphics[width=0.3\textwidth]{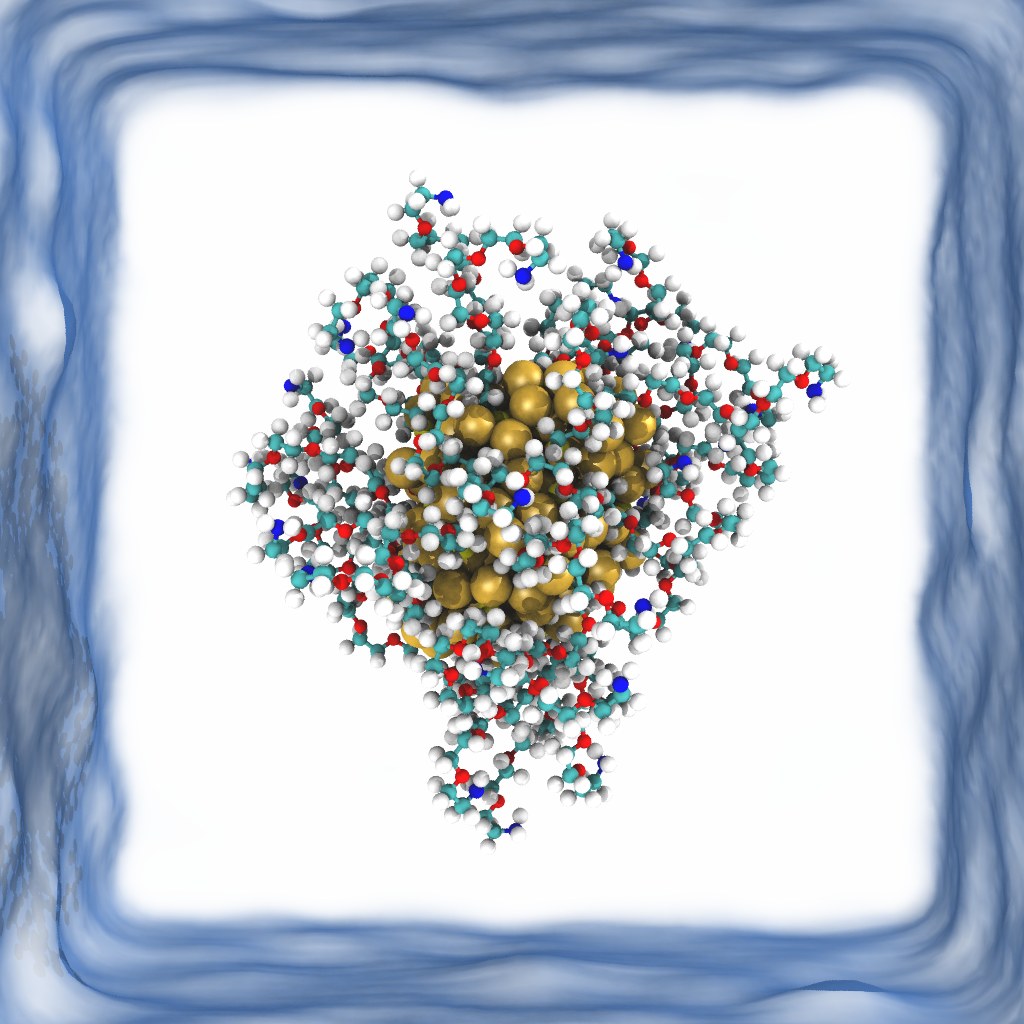}
		}
		\par\end{centering}
	\caption{(a) Overview of initial configuration for $N_\mathrm{PEG} = 32$ (waterbox dimensions reduced for this illustration), (b) after annealing to \SI{0}{K}.}
	\label{fig:32PEGwaterbox-initial-and-final}
\end{figure}

This optimized system was then annealed by first heating it up to and keeping it at \SI{1000}{K} for \SI{400}{ps} followed by a step-wise cooling of \SI{100}{K} per \SI{100}{ps} down to \SI{0}{K}. The difference between initial and final structure of the system is illustrated in Fig.~\ref{fig:32PEGwaterbox-initial-and-final}.

\section{Results and discussion}

\subsection{Coating thickness}
To evaluate theoretically the conformation of polymers attached to a surface and the resulting thickness of polymer coatings it is customary to apply the framework detailed by de Gennes \citep{DeGennes1980}. Two regimes are defined, the first being the low-density regime where the surface ligand density is so low that the polymers are essentially isolated and free to coil around themselves. Due to the semi-spherical shape they attain, this is know as the "mushroom" regime and is realized when the distance $D$ between ligand grafting points is larger than the Flory radius $R_\mathrm{F}$ of the polymer given by \citep{Flory1953}
\begin{equation}
	R_\mathrm{F} = aN^{3/5},
	\label{eq:Flory}
\end{equation}
where $a$ is the monomer length and $N$ is the number of monomers in the polymer. In the high-density regime, defined as when $D < R_\mathrm{F}$, the interaction between the closely spaced polymers cause them to attain a more linear shape stretching up from the surface and is therefore referred to as the "brush" regime. The resulting coating thickness is given by \citep{DeGennes1980}
\begin{equation}
L = Na\left( \frac{a}{D}\right) ^{2/3}.
\label{eq:Lbrush}
\end{equation}

The distance between ligands $D$ can be calculated by approximating the AuNP as a spherical particle with diameter $d = \SI{1.4}{nm}$ and surface area $S = 4 \pi ( d/2 )^2$. Assuming the average surface area per ligand $A = S/N_\mathrm{PEG}$ as circular, $D$ is then the diameter of this circle, 
\begin{equation}
	D = 2 \sqrt{\frac{S}{\pi N_\mathrm{PEG}}} = 2 \frac{d}{\sqrt{N_\mathrm{PEG}}}
\end{equation}
giving $D_{32} = \SI{5.0}{\AA}$ and $D_{60} = \SI{3.6}{\AA}$, for $N_\mathrm{PEG} = 32$ and 60, respectively --- significantly below the Flory radius for the PEG considered in this paper which is $R_\mathrm{F} = \SI{9.2}{\AA}$ using $N = 5$ and $a = \SI{3.5}{\AA}$ \citep{Rahme2013}.

\begin{figure}
	\begin{centering}
		\includegraphics[width=0.90\columnwidth]{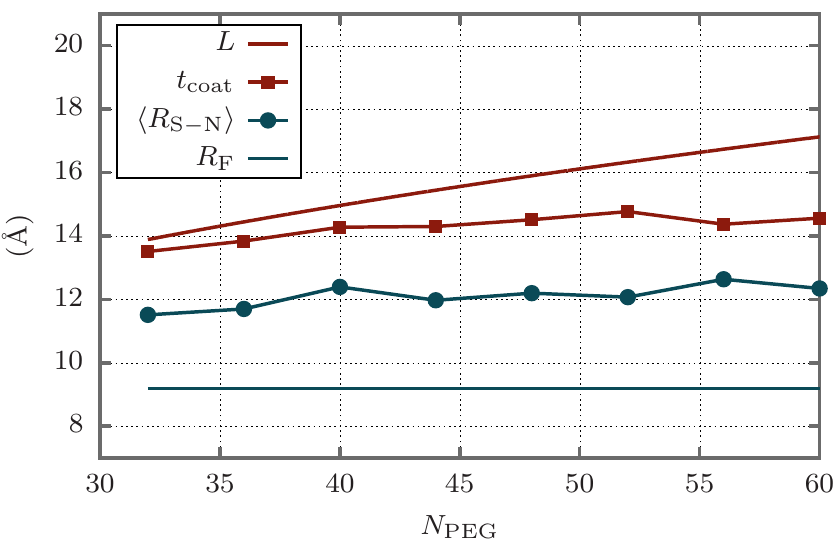}
		\par\end{centering}
	\caption{\label{fig:thickness-endtoendVsNumpeg}Theoretical brush regime thickness $L$ (Eq.~\eqref{eq:Lbrush}, see text) as well as the thickness of the coating $t_{\mathrm{coat}}$ as a function of the number of attached PEG molecules $N_{\mathrm{PEG}}$ calculated from the average position of the sulfur atoms to the distance which contains 97\% of the coating atoms. Also shown are the average end-to-end distance $\left\langle R_\mathrm{S-N}\right\rangle$ for each value of $N_{\mathrm{PEG}}$ and the Flory radius $R_\mathrm{F}$ (Eq.~\eqref{eq:Flory}).}
\end{figure}

Using Eq.~\eqref{eq:Lbrush}, we can obtain theoretical estimates of the brush thicknesses $L_{32} = \SI{13.9}{\AA}$ and $L_{60} = \SI{17.1}{\AA}$ for $N_\mathrm{PEG} = 32$ and 60. The coating thickness $t_\mathrm{coat}$ measured from the simulations is plotted in Fig.~\ref{fig:thickness-endtoendVsNumpeg} together with the theoretical estimates given by Eq.~\eqref{eq:Lbrush}. $t_\mathrm{coat}$ was calculated as the thickness which contained 97\% of the coating atoms measured from the average position of the sulfur atoms. As can be seen in the figure, $t_\mathrm{coat} < L$ for all values of $N_\mathrm{PEG}$ but most interestingly $t_\mathrm{coat}$ is almost independent of $N_\mathrm{PEG}$ and therefore of the ligand surface density, which increases from \SIrange{5.2}{9.7}{nm^{-2}} as $N_\mathrm{PEG}$ increases from 32 to 60 --- compare with the values listed in Table~\ref{tab:pegCoatings}. This is most likely a consequence of the strongly curved surface of the \SI{1.4}{nm} AuNP, as discussed below.

In Table~\ref{tab:pegCoatings}, a number of PEG surface ligand density measurements are presented as a function of the size of the AuNP and the weight of the PEG molecules. The densities reported are mostly lower than what we found, which we ascribe to the rather big differences in NP size and PEG weight. Additionally, in our "synthesis", the PEG molecules started off as linear, which allows for a denser packing than the coiled structure PEGs have in suspension during synthesis in experimental conditions.

To compare with the mushroom regime, the average end-to-end distance $\left\langle R_\mathrm{S-N}\right\rangle$ for each value of $N_\mathrm{PEG}$ is plotted in Fig.~\ref{fig:thickness-endtoendVsNumpeg} together $R_\mathrm{F} = aN^{3/5}$. The fact that the end-to-end distance is larger than the Flory radius, but not as large as the coating thickness, indicates that the PEG coating considered here is in a mixed state between mushroom and brush.

As seen from Fig.~\ref{fig:thickness-endtoendVsNumpeg}, there is a systematic discrepancy between the measured coating thickness $t_\mathrm{coat}$ and the theoretical brush thickness $L$ as given by Eq.~\eqref{eq:Lbrush}. We do not believe this to be due to the short chain length; it was shown by Zimmt et al. that a Gaussian spatial distribution, as assumed for the framework introduced by de Gennes, is still a good description for chains as short as three segments \citep{Zimmt1988}.

Instead, we believe the most important factor to be the shape of the surface. Although the mushroom and brush regimes are defined for flat surfaces, they are routinely employed for coatings on NPs, but often these have sizes larger than \SI{30}{nm}. In the present study, the highly curved surface of a \SI{1.4}{nm} NP will significantly reduce the steric repulsion between PEG chains allowing for more coiled chains than for a flat surface of similar ligand surface density, as found by Walkey et al. \citep{Walkey2012}. In order to apply the mushroom/brush formalism to small NPs the equations should be modified to allow for a curved rather than flat surface. This, however, is beyond the scope of this paper.

\begin{table}
	\centering
	\caption{Summary of PEG coating surface ligand densities $\theta$ for various AuNP sizes $d$ and PEG weights $W$.}
	\label{tab:pegCoatings}
	\begin{tabular}{
			S[table-format=3.2] 
			S[table-format=3.2] 
			S[table-format=3.2] 
			l } \toprule[1pt]
		
		\multicolumn{1}{c}		{$d$ 	(nm)}	
		& \multicolumn{1}{c}	{$W$ (kDa)}	
		& \multicolumn{1}{c}	{$\theta$ 	(\SI{}{nm^{-2}}) } 
		& Ref \\ \midrule

30	&   2		&  9.2 	& 	\citep{Smith2015}	\\	\addlinespace[1em]

30	&   2.1		&  3.93 & 	\citep{Rahme2013}	\\
70	&   		&  0.31	& 						\\
150	&   		&  0.31	&						\\	\addlinespace[1em]

30	&   2		&  2.30 &	\citep{Hansen2015}	\\
  	&   5		&  0.92	&						\\
  	&   10		&  0.33	&						\\
  	&   20		&  0.28	&						\\	\addlinespace[1em]
		
60	&   1		&  1.55 & 	\citep{Tsai2011}	\\
 	&   5		&  0.17	&						\\
  	&   20		&  0.025&						\\  \addlinespace[1em]

2.8	& 	5		&  2.9	& 	\citep{Wuelfing1998}\\	\addlinespace[1em]
		  	 
1.5	&   0.27	&   \SIrange{5.2}{9.7}{}&	This work\\	\bottomrule[1pt]
	\end{tabular}
\end{table}


\subsection{Density distribution}
\begin{figure}[h]
	\begin{centering}
		\subfloat[] {
			\includegraphics[width=0.90\columnwidth]{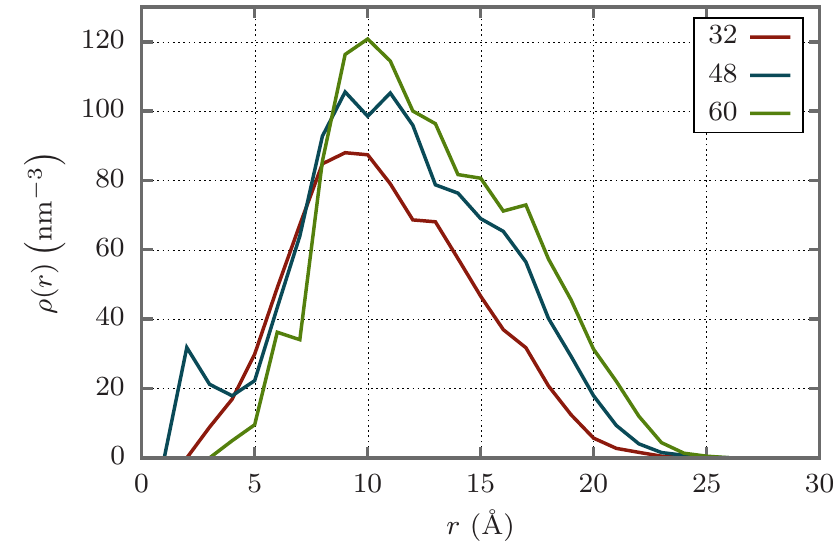}
			\label{fig:totalDensity}
		} \qquad
		\subfloat[] {
			\includegraphics[width=0.90\columnwidth]{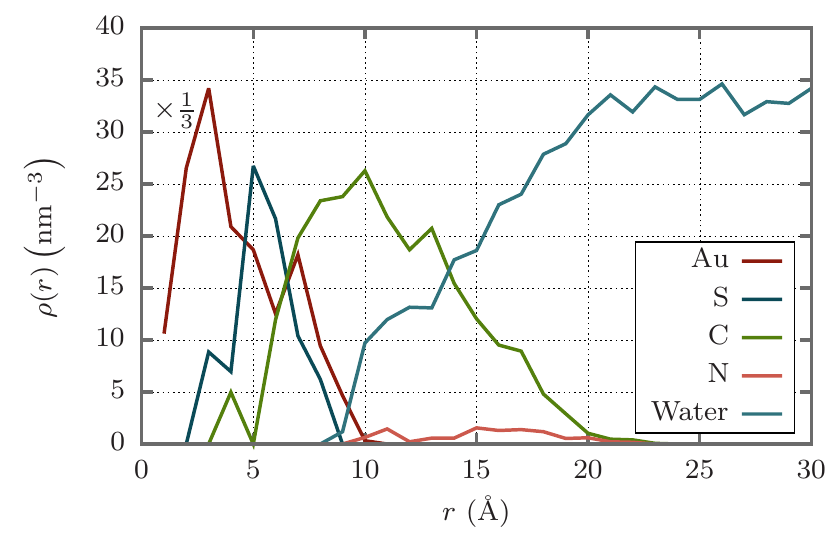}
			\label{fig:elementDensity32}
		} \qquad
		\subfloat[] {
			\includegraphics[width=0.90\columnwidth]{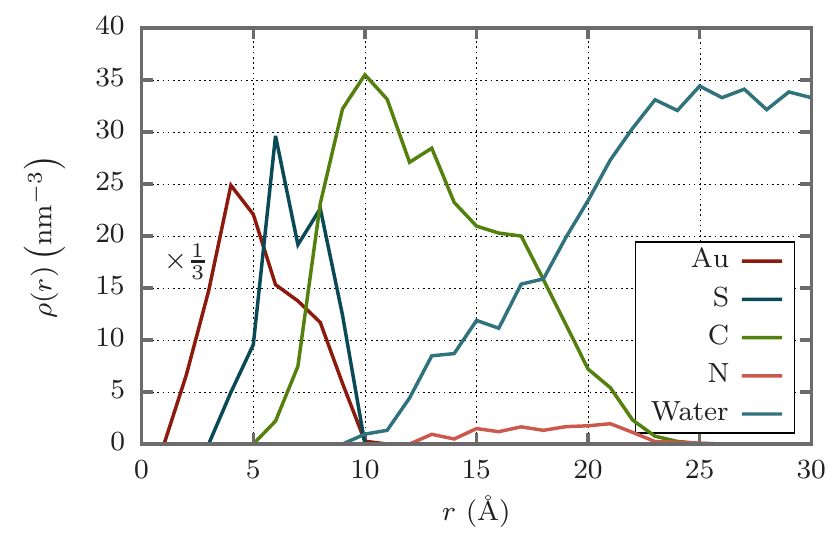}
			\label{fig:elementDensity60}
		}
		\par\end{centering}
	\caption{Density distribution function after annealing to \SI{310}{K}. (a) Total density distribution for $N_\mathrm{PEG} = 32$, 48, and 60. (b) and (c) show partial density distribution function for the elements of the coating S, C, and N, as well as for water, for $N_\mathrm{PEG} = 32$ and $N_\mathrm{PEG} = 60$, respectively. Note that the Au line has been rescaled by 1/3.}
	\label{fig:Density-plots}
\end{figure}

The density distribution was calculated for each of the simulated NPs by dividing the radial distribution, found by counting the number of atoms belonging to the ligands in concentric shells around the center of mass of the system, by the volume of each shell $V_{\mathrm{s}}=4\pi r^{2}\mathrm{d}r$, where $\mathrm{d}r=\SI{1.0}{A}$ is the shell thickness. Figure~\ref{fig:Density-plots} shows the density distribution for $T = \SI{310}{K}$ for different values of $N_\mathrm{PEG}$.

From Fig.~\subref*{fig:totalDensity}, it is seen that the total density increases with $N_\mathrm{PEG}$, as expected. Comparing Fig.~\subref*{fig:totalDensity} with Figs. \subref*{fig:elementDensity32} and~\subref{fig:elementDensity60}, which show the partial densities, it can be seen, that the peak total density coincides with the maximum extent of the sulfur atoms, around $r = \SI{10}{\AA}$, which are located at slightly larger distances from the center of mass for higher values $N_\mathrm{PEG}$. This can be ascribed to the repulsion between sulfur atoms increasing for higher surface densities thereby reducing the the intercalation. The position of the peak density is therefore only indirectly dependent on $N_\mathrm{PEG}$.

From Figs. \subref*{fig:elementDensity32} and~\subref{fig:elementDensity60} we can evaluate the amount of water penetration into the coating. The most pronounced difference between the two cases is the water density close to the gold surface. For $N_\mathrm{PEG} = 32$, the density rises steeply to about $\SI{10}{nm^{-3}}$ around $r = \SI{10}{\AA}$ before a more moderate increase until it plateaus off around $\SI{35}{nm^{-3}}$, which is close to the density of water at \SI{310}{K}. In contrast for $N_\mathrm{PEG} = 60$ the water density increases steadily and does not reach a density of $\SI{10}{nm^{-3}}$ until around $r = \SI{14}{\AA}$. 

Hydroxyl yield under radiation, and therefore potential for radiosensitivity, could be dependent on the density of the PEG coating, as shown by Gilles et al. \citep{Gilles2014}. PEG is a hydrophilic polymer and it has been proposed that the way in which PEG works to protect the NP from clearance from the bloodstream is related to its ability to trap water molecules \citep{Stolnik1995}. It is therefore interesting to monitor both the density of the coating as well as the degree of water penetration into the coating.

It should be noted that due to the partial reconstruction of the AuNP during the simulations, some gold atoms are located farther from the center of mass than some of the other atoms, which is why it appears that a significant number of carbon atoms as well as some water molecules have penetrated the Au surface. This is not the case, but is a combination of the reconstruction of the Au core and how the density distribution is calculated.

\section{Conclusion}
PEGylated AuNPs of core size \SI{1.4}{nm} were simulated using classical molecular dynamics. The AuNPs were coated with between 32 and 60 thiolated PEG-amine ligands\linebreak($\mathrm{S-PEG_5-NH_2}$) and the thickness and density distribution of the NPs was presented.

It was shown that the mushroom/brush regime usually employed to analyze the thickness and surface density of PEGylated NPs should be used with caution for small, highly curved NPs. The thickness of the coating was found to be only weakly dependent on the ligand surface density as was the end-to-end distance of the ligands. It was shown that the water penetration into the PEG coating was increased for lower number of attached ligands --- which can have an important effect on the hydroxyl yield when the NP is irradiated.

Finally it should be noted that there are many other properties of NPs that may change as a function of coating for example they may influence the pH of the host medium and the transport through the medium. Similarly the size, shape, and composition of the NP core will be important. Due to the computational cost of approximately \SI{4500}{CPU-hours} per simulation, this is planned to be explored in future studies using the methodology developed in this paper.

\bigskip{}

The research leading to these results has received funding from the European Union Seventh Framework Programme PEOPLE - 2013 - ITN - ARGENT project under grant agreement number 608163.


\bibliographystyle{epj}
\bibliography{library}

\end{document}